\documentclass{iopart}
\usepackage{graphicx,amsfonts,amsbsy,amssymb,amsthm}

\expandafter\let\csname equation*\endcsname\relax
\expandafter\let\csname endequation*\endcsname\relax
\usepackage{amsmath}

\newcommand{\be}{\begin{equation}}
\newcommand{\ee}{  \end{equation}}
\newcommand{\ba}{\begin{eqnarray}}
\newcommand{\ea}{  \end{eqnarray}}

\def\={\;=\;} \def\+{\,+\,}

\def\cl{^{(cl)}}
\def\ccl{C^{(cl)}}

\begin{document}

\title[Delay-time distribution in the scattering of time-narrow wave packets
  (II)]{Delay-time distribution in the scattering of time-narrow wave packets
  (II) - Quantum Graphs}

\author{Uzy Smilansky$^{1}$ and  Holger Schanz$^{2}$ }

\address {$^{1}$Department of Physics of Complex Systems,
 Weizmann Institute of Science, Rehovot 7610001, Israel.}
\address {$^{2}$Institute of Mechanical Engineering, University
  of Applied Sciences Magdeburg-Stendal, D-39114 Magdeburg, Germany.}       

 \date{\today}

\begin{abstract}
 We apply the framework developed in the preceding paper in this series
 \cite{Smi17} to compute the time-delay distribution in the scattering of ultra
 short RF pulses on complex networks of transmission lines which are modeled
 by metric (quantum) graphs. We consider wave packets which are centered at
 high quantum number and comprise many energy levels. In the limit of pulses
 of very short duration we compute upper and lower bounds to the actual time
 delay distribution of the radiation emerging from the network using a
 simplified problem where time is replaced by the discrete count of
 vertex-scattering events.  The classical limit of the time-delay distribution
 is also discussed and we show that for finite networks it decays
 exponentially, with a decay constant which depends on the graph connectivity
 and the distribution of its edge lengths. We illustrate and apply our theory
 to a simple model graph where an algebraic decay of the quantum time delay
 distribution is established.
\end{abstract}
\maketitle

\section{Introduction}
\subsection {Motivations}
When an ultra-short pulse of radiation is scattered on a complex medium, the
emerging radiation pulse is broadened in time and the pulse shape reflects the
distribution of time-delays induced in the scattering process. This
distribution can be intuitively explained as due to the existence of a large
number of paths of varying lengths through which the radiation can traverse
the scatterer. Recently, novel methods to produce ultra-short light pulses
were introduced. They opened a new horizon for experiments where the
distribution of delay-times induced by scattering from complex targets can be
measured, with interesting and surprising results, see e.g.,
\cite{E+08,L+14,H+16}. The ultra-short pulses are realized as broad-band
coherent wave packets, which are presently available only for electromagnetic
radiation, but not yet for sub-atomic particles such as e.g.,
electrons. However, work towards this end has already begun
\cite{MPG-LMU15}. These developments emphasize the need for theoretical tools
to aid planning of new experiments and interpret the measured results.  

The preceding paper in this series \cite{Smi17} presented a general
theoretical framework for the computation of the delay-time distribution in
scattering of short radiation pulses on complex targets. The ingredients which
are needed in this theory are the scattering matrix $S(k)$ where $k$ is the
wave-number, the pulse (wave-packet) envelope $\omega(k)$ and the dispersion
relation $E(k)$. For scattering of electromagnetic radiation the latter is
$E(k)=ck$ where $c$ denotes the velocity of light. In this case it is
convenient to express the time by the optical path-length $s=tc$. The general
expression for the delay-time distribution is then given by
 \begin{equation}
\hspace{-15mm}
P_{{f},{i}}(s) =  \frac{1}{2\pi}\  \left|\int_{0}^{\infty} {\rm d}k   \ \omega(k) S_{f, i}(k) e^{-iks}\right|^2
\label{gpoftem1}
\end{equation}
if the delay is measured for pulses impinging in channel $i$ and detected in
channel $f$.

In the present paper, we apply this general formalism to scattering on
  quanum graphs \cite {kottos1,kottos2,gnutzmann,berkokuch,rami}. We do so for
  two reasons: First, quantum graphs are known as a successful paradigm for
  scattering from complex targets while at same time they are analytically and
  numerically much more tractable. For example we will present in this paper a
  full analytic solution of a model which contains some essential ingredients
  for complex targets such as an exponetially increasing number of scattering
  trajectories and relevant quantum interferences between them. Thus, studying
  quantum graphs in the present context might reveal typical features which
  are difficult to decipher in more realistic systems. Second, quantum graphs
  are very good models for the scattering of radio frequency (RF) signals in
  networks of wave-guides. As a matter of fact, experiments on the delay-time
  distribution in such systems are presently performed in Maryland and Warsaw
  \cite{Anlage, Sirko}.

\subsection{Outline}
In the following Section \ref{sec:qgr} the necessary definitions and tools
from the theory of quantum graphs will be provided.
Then in Section \ref{sec:scatgr} this theory will be extended to scattering on
graphs and an explicit formula for the scattering matrix $S_{f,i}$ will be
discussed.
In Section \ref{RF} we apply this formula to Eq.~(\ref{gpoftem1}) and derive
on this basis approximate expressions for the delay time distribution in the
case of broad envelope functions $\omega(k)$ corresponding to wave packets
narrow in time.

Section \ref{diagonal} is devoted to the clasical analogue of the delay
  time distribution. In particular we show that for finite and connected
  graphs the classical delay distribution decays exponentially for long times
  and calculate the decay exponent.  The classical delay distribution provides
  both, a simple short-time approximation to the fully coherent expression
  (\ref{gpoftem1}) and a reference result which allows to highlight quantum
  interference contributions to (\ref{gpoftem1}) for longer times. Moreover,
  as in the above mentioned experiments with RF radiation some decoherence
  cannot be avoided, a satisfactory theory might involve a crossover between
  our results for coherent and incoherent time delay.

  In the final Section \ref{sec:tjunction} we apply all our results to a
  simple model system which consists of two edges and a single scattering
  channel. For this model we can also confirm the results of Section \ref{RF}
  by an independent calculation based on the distribution of narrow scattering
  resonances.

\subsection {Quantum graphs in a nut-shell}\label{sec:qgr}
A graph $\mathcal{G}(\mathcal{V,E})$ consists of a finite set of vertices
$\mathcal{V} ,\ |\mathcal{V}|=V$ and edges $\mathcal{E}
,\ |\mathcal{E}|=E$. It will be assumed that $\mathcal{G}$ is connected and
simple (no parallel edges and no self connecting loops). The connectivity of
$\mathcal{G}$ is encoded in the $V\times V$ adjacency matrix $A$: $A_{u,v} =
1$ if the vertices $u,v\in \mathcal{V}$ are connected and $A_{u,v} = 0$
otherwise. The set of edges connected to the vertex $v$ is denoted by
$\mathcal{S}(v)$. The degree of the vertex $v$ is $d_v =
|\mathcal{S}(v)|$. When $A_{u,v} = 1$, the connecting edge $e = (u,v)$ will be
endowed with two directions $\varpi =\pm $, the positive direction is chosen
to point from the lower indexed vertex to the higher. A pair $d= (e,\varpi)$
is a directed edge. The set of all directed edges will be denoted by
$\mathcal{D}$ and $D=|\mathcal {D}|=2E$ is its size. The reverse of
$d$ is denoted by ${\hat d} = (e,-\varpi)$.  When $d$ is a directed edge
pointing from vertex $u$ (the origin of $d$) to $v$ (the terminus of $d$) we
write $u= o(d)$ and $v=t(d)$, respectively.

 An alternative way to describe the connectivity of $\mathcal{G}$ is in terms
 of the edge adjacency matrix of dimension $D\times D$ :
\begin{equation}
B_{d,d'}=\delta_{o(d),t(d')} \ , \ \ d,d'\in \mathcal{D} \ .
\end{equation}

\vspace {2mm}

The metric endowed to the graph is the natural one-dimensional Euclidian
metric on every edge. The length of an edge $e$ is denoted by $L_e$ and
$\mathcal{L}= [ L_e ]_{e=1}^E$ is the set of these edge lengths. The edge
lengths $L_{e}$ are assumed to be rationally independent.  A graph is compact
when all the edge lengths $L_e$ are finite.  The lengths of the directed edges
$d$ and its reverse $\hat d$ are equal. Denote by $x_e$ the coordinate of a
point on the edge $e$, measured from the vertex with the smaller index and
$0\le x_e \le L_e$. A function $F: x\in \mathcal{G} \rightarrow \mathbb{R}$ is
given in terms the functions $[f_e (x_e)]_{e=1}^E$ so that if $x\in e,
\ F(x)=f_e(x)$.  The action of the Laplacian on $F(x)$ for $x\in e$ is
$\Delta_{\mathcal{G}}F = -\frac{\partial ^2 f_e }{\partial x_e ^2}$ and the
domain of the Laplacian is $f_e ({x_e})\in \mathcal{C}^2(0,L_e), \ \forall
e$. Assume $\mathcal{G}$ is compact. Then, the Laplacian is self-adjoint if it
acts on a restricted space of functions $F(x)$ which satisfy appropriate
boundary conditions. Frequently used boundary conditions are the Neumann
conditions which require the function $F$ to be continuous at all the
vertices, and for every vertex, $\sum_{e\in \mathcal{S}(v)}\frac{\partial
  f_e(x_e)}{\partial x_e}|_{v} =0 $, where the derivatives at $v$ are taken in
the direction which points away from $v$. The most general prescriptions for
boundary conditions were first introduced and discussed in
\cite{Schrader}. The time dependent wave equation (with time $s/c$) is
\begin{equation}
\frac{\partial^2\ } {\partial s^2} F(x,s)= \Delta_{\mathcal{G}}F(x,s)
\end{equation}
with the boundary conditions specified above which must be satisfied for all
$s$. The stationary equation
\begin{equation}
 \Delta_{\mathcal{G}}F(x,k) = k^2 F(x,k)\ ,
\end{equation}
can be solved only for a discrete, yet infinite set of wave numbers $[k_n]_{n \in \mathbb{Z}}$ which is the spectrum of the stationary wave equation.

A useful method of computing the spectrum of the graph Laplacian is based on
the following decomposition of the wave function. Consider the functions
$f_e(x_e) = a_d e^{i\varpi k x_e} + a_{\hat d}e^{i\varpi k (L_e- x_e)}$, where
$d=(e,\varpi)$ and $a_d, a_{\hat d}$ are arbitrary complex numbers. These
functions are the general solutions of $-\frac{\partial ^2 f_e }{\partial x_e
  ^2} =k^2f_e(x_e)$ on all the edges. The constants should be computed so that
$F(x)$ satisfies the boundary conditions at all the vertices. Consider all the
edges which are connected to a vertex $v : \ e\in \mathcal{S}(v)$. For Neumann
boundary conditions the continuity of the graph wave function at the vertex
$v$ imposes $d_v-1$ independent requirements on the coefficients
$a_d$. Namely, $f_e|_v = f_{e'}|_v \forall e \ne e' \in {\mathcal{S}(v)}$,
where $f_e|_v$ denotes the value of $f_e$ at the vertex $v$, where $x_e=0$ or
$x_e=L_e$ depending on the orientation of $d$.  Again for Neumann boundary
conditions, another relation among the $a_d$ is imposed by the requirement
that the sum of the outgoing derivatives of the $f_e$ at the vertex
vanishes. Therefore there are $d_v$ linear equations which the $2d_v$
coefficients must satisfy. Hence, if one denotes the set of directed edges
which point towards $v$ by $\mathcal{S}^{-}(v)$ and the complementary set of
outgoing directed edges by $\mathcal{S}^{+}(v)$, then the boundary conditions
at $v$ provide a linear relation between the two subsets of coefficients:
\begin{equation}
\hspace{-10mm} a_d =\sum_{d'\in \mathcal{S}^{-}(v) } \sigma^{(v)}_{d,d'} a_{d'} ,\ \ \  \forall d\in \mathcal{S}^{+}(v)\ , \ \ \ {\rm with}\ \  \sigma^{(v)}_{d,d'}=\frac{2}{d_v}-\delta_{d,{\hat d}'}\ .
\end{equation}
The symmetric and unitary matrix $\sigma^{(v)}$ of dimension $d_v$ is the
vertex scattering matrix corresponding to Neumann boundary condition at the
vertex. Other boundary conditions yield different vertex scattering matrices,
and their unitarity is due to the fact that the underlying graph Laplacian is
self adjoint. Using the vertex scattering matrices for all the vertices on the
graph, one can construct a $D \times D$ unitary matrix
\begin{equation}
U_{d,d'}(k) = \delta_{o(d),t(d')}e^{ikL_d} \sigma^{(o(d))}_{d,d'}\ , \ d,d'\in \mathcal{D} \ ,
\label{defU}
\end{equation}
which acts on the $D$ dimensional space of complex coefficients $a_d$. It then
follows \cite{kottos1,kottos2} that the spectrum of the graph Laplacian is
obtained for values of $k$ which satisfy the secular equation
\begin {equation}
\det [I^{(D)}- U (k)] \ = \ 0  \ ,
\end {equation}
where $I^{(D)}$ is the unit matrix in dimension $D$. The unitarity of $U(k)$ for
real $k$ implies that all the eigenvalues of $U(k)$ are on the unit circle. As
$k$ varies, eigenvalues cross the real axis, where the secular equation is
satisfied. Therefore the $k$ spectrum is real.

$U(k)$ is referred to as the graph evolution operator in the quantum chaos
literature.  Its matrix elements provide the amplitudes for scattering from an
edge $d$ directed to a vertex $v$, to an edge $d'$ directed away from
$v$. Their absolute squares can be interpreted as the probabilities that a
classical particle confined to the graph and moving on the edge $d$ toward the
vertex $v$ is scattered to the edge $d'$ and moves away from it. Due to the
unitarity of $U$ the $D\times D$ matrix $M$
\begin {equation}
M_{d,d'} = |U_{d,d'}|^2 \ ,\  \ d,d'\in \mathcal{D} \ ,
\label{defM}
\end {equation}
does not depend on $k$ and is double Markovian: $ \sum_{d}M_{d,d'} =
\sum_{d'}M_{d,d'}=1 $. The transition probability matrix $M$ allows to define a
random walk on the graph. For the graphs considered here, $M$ satisfies the
conditions of the Frobenius-Perron theorem and therefore the largest
eigenvalue of $M$ is $1$ and it is single. Suppose that at time $0$ the
probability distribution to find the walker on the directed edge $d$ is given
by the vector $p_{d}(t=0), \ \forall d\in \mathcal {D}$. Then, at integer
time $t>0$ the distribution will be $p(t)=M^tp(0)$ and converges to
equidistribution for large $t$ independently of the initial probability
distribution. In other words, the classical evolution on the graph is ergodic.
(Note that we will use the symbol $t$ for a discretized {\em topological}
time while the continuous physical time is measured in terms of the path
length $s$ as in Eq.~(\ref{gpoftem1}).)

\subsection {Scattering on quantum graphs}\label{sec:scatgr}
So far we discussed the wave equation and its classical limit on a compact
graph. To turn this graph into a scattering system, we choose a subset of
vertices $\mathcal{H}\in \mathcal{V}$, and at every vertex $h\in \mathcal{H}$
we add a semi-infinite edge (lead). $H=|\mathcal{H}|$ is the number of
leads. The directed edges on the lead attached to vertex $h$ are denoted by
$h^{(+)}$ which points away from the vertex $h$ and $h^{(-)}$ which points
towards it. The Laplacian is extended to the leads in a natural way, and the
boundary conditions at the vertices $h\in \mathcal{H}$ are modified by
replacing $d_v$ by $d_h=d_v+1$. Measuring distances from the vertex $h$
outwards, the functions which are allowed on the lead take the form $f_h(x )=
a_{h^{(-)}} e^{-ikx } + a_{h^{(+)}} e^{ikx}$. The spectrum of the Laplacian
for a scattering graph is continuous and covers the entire real line, possibly
with a discrete set of embedded eigenvalues (See e.g., \cite{rami}).

Consider the matrix
\begin{equation}
 W_{d,d'} = \delta_{o(d),t(d')}e^{ikL_d}{\tilde \sigma}^{(o(d))}_{d,d'}\ , \ \ d,d'\in \mathcal{D} \ ,
\end{equation}
where ${\tilde \sigma}^{(u)}$ for $u\in \mathcal{H}$ are the vertex scattering
matrices which are modified as explained above, and for $u$ in the complement
of $\mathcal{H}$, they take the values of the vertex scattering matrices for
the compact graph. Note that $W(k)$ is a $D\times D$ matrix, and its entries
are indexed by the labels of the directed edges in the compact part of the
graph, in the same way as the original matrix $U(k)$ of
Eq.~(\ref{defU}). However, unlike $U(k)$, $W(k)$ is not unitary, because some
of its building blocks, namely the vertex scattering matrices ${\tilde \sigma}
^{(h)}, h\in \mathcal{H}$ are not unitary when they are restricted to the
directed edges in the compact part of the graph.

The analogue of $M$ defined in (\ref{defM}) for the non-compact graph is
\begin{equation}
{\tilde M}_{d,d'} = |W_{d,d'}|^2=  \delta_{o(d),t(d')}\left|{\tilde \sigma}^{(o(d))}_{d,d'}\right|^2\ .
\label{clasprob}
\end{equation}
It is independent of $k$ and sub-Markovian since for $o(d)\in\mathcal{H}$ the
sums $\sum_{d'}{\tilde M}_{d,d'}$ and $\sum_{d'}{\tilde M}_{d',d}$ are
strictly less than $1$. The Perron-Frobenius theorem guarantees that the
spectrum of $\tilde M$ is confined to the interior of the unit circle. For a
random walker whose evolution is dictated by $ \tilde M$, the probability to
stay inside the compact part of the graph approaches zero after sufficiently
long time. This is due to the walks which escape to the leads and never
return.

Consider now a solution of the stationary wave equation for a given $k$
subject to the condition that the wave function on the leads has the form $f_h
(x_h) = a_{h^{(-)}}e^{-ikx_h}+a_{h^{(+)}}e^{+ikx_h}$. The scattering matrix for
a non compact graph is a unitary matrix of dimension $H$ which provides the
vector of "outgoing amplitudes" ${\bf a}^{(+)}=\{a_{h^{(+)}}\}_{h\in H}$ in
terms of the vector of "incoming amplitudes" ${\bf
  a}^{(-)}=\{a_{h^{(-)}}\}_{h\in H}$. It follows from the linearity of the
wave equation that
\begin{equation}
{\bf a}^{(+)}=S(k){\bf a}^{(-)}\ .
\end{equation}
The explicit expression for $S(k)$ was derived in \cite{kottosscatter,
rami} and will be quoted here without proof:
\begin{align}
\label{defS}
S_{h,h'}(k) &=\delta_{h,\hat h'} \rho_{h'} +\sum_{d,d'} \tau_{h,d}
  \left \{\sum_{n=0}^{\infty}[W^n(k)]_{ d,d'}\right \}\ e^{ikL_{d'} }\tau_{d',h'} \nonumber \\
&=\delta_{h,\hat h'} \rho_{h'} +\sum_{d,d'} \tau_{h,d}
   \left[I^{(D)}- W(k)\right]^{-1}_{d,d'}\ e^{ikL_{d'} }\tau_{d',h'}\ .
\end{align}
Here, $\rho_h={\tilde \sigma}^{(h)}_{{\hat h},h}$ is the back reflection
amplitude, $\tau_{d',h'}={\tilde \sigma}^{(h')}_{d',h'}$ is the transmission
amplitude from the lead $h'$ to the edge $d'$ in the compact part of the
graph, and $\tau_{h,d}$ is the transmission amplitude from an edge $d$ in the
compact graph to a lead $h$. The first line in (\ref{defS}) expresses the fact
that scattering proceeds by either reflecting from the incoming lead back to
itself (the term outside the sum), or by penetrating to the compact part and
scattering inside it several times before emerging outside. The contribution
of the scattering process in the compact graph is provided by the expression
in curly brackets. It can be rewritten as
\begin{eqnarray}
\hspace{-5mm}
 \sum_{d,d'} \tau_{h,d}\left \{\sum_{n=0}^{\infty}[W^n(k)]_{d,d'}\right \}\ e^{ikL_{d'} }\tau_{d',h'}
= \sum_{n=0}^{\infty} \sum_{\alpha\in {\mathcal{A}}_{h,h'}^{(n)}} A^{(n)}_{\alpha}e^{ikl_{\alpha}}  .
\label{sfrompaths}
\end{eqnarray}
Here, $n$ counts the number of vertices on a path $\alpha$ connecting the
entrance and exit vertices $h'$ and $h$.  ${\mathcal{A}}^{(n)}_{h,h'}$ is the
set of all the paths crossing $n$ vertices which start on $h'$ and end at $h$
after traversing $n+1$ directed edges $(d_0,d_1,\cdots,d_n), d_j\in
\mathcal{D}$ with $o(d_0)=h',t(d_n)=h$. Each path is of length $l_{\alpha} =
\sum_{j=0}^n L_{d_j}$.  The term $n=0$ occurs only when $h'$ and $h$ are
neighbors on $\mathcal{G}$. Then $\alpha$ is the directed edge $d$ connecting
$h'$ to $h$, $\mathcal{A}_{h,h'}^{(0)}$ consists of the single bond $d$,
$A^{(0)}=\delta_{d,d'}\tau_{h,d}\tau_{d,h'}$ and $l_{\alpha}=L_d$. For $n\ge1$ the
amplitudes $A_{\alpha}^{(n)}$ can be written as
\begin{equation}
A_{\alpha}^{(n)}=\tau_{h,d_n}\left [ \prod_{j=1}^n
  \sigma^{(o(d_j))}_{d_j,d_{j-1}}\right ]\tau_{d',h'}\ .
\label{aofalpha}
\end{equation}
  The series in the first line of (\ref{defS}) converges to the expression in
  the second line for any real $k$ because $W(k)$ is sub-unitary. The explicit
  form of the $S(k)$ matrix provided in Eq.~(\ref{defS}) will be used in the
  next sections. An alternative expression for $S(k)$ which will not be used
  here can be found in \cite{kottos2}.

\section{Scattering of wave-packets and the delay-time distribution }
\label{RF}

Given a graph with leads to infinity as defined above, we consider a
particular solution of the stationary wave equation with wave number $k$,
where the wave function consists of an incoming wave with unit amplitude in a
single lead $h'$ but outgoing waves in all the leads. Limiting our attention
to a specific lead $h$ the wave function has the form
\begin{equation}
\hspace{-10mm}f_h(x_h) = \delta_{h,h'}e^{-ikx_h}+a_{h{^{(+)}}} e^{ikx_h} = \delta_{h,h'}e^{-ikx_h} +S_{h,h'}(k)e^{ikx_h}
\end{equation}
The last equality follows from the definition of the scattering matrix. A
time-dependent solution describing the propagation of a wave packet is
obtained by a superposition of functions $f_h(x_h)$ with an envelope function
$\omega (k)$. As in \cite{Smi17} $\omega(k)$ is positive and normalized by
$\int_0^{\infty}\omega^2(k){\rm d}k = 1$. Assuming a linear dispersion relation
(such as e.g. for electromagnetic waves in transmission lines), the intensity
of the outgoing wave function in the position $x_h=0$ at time $s/c$ is
\begin{equation}\label{ps}
P_{h,h'}(s)=\frac{1}{2\pi} \left | \int_0^{\infty} \omega(k) S_{h,h'}(k) e^{-iks} {\rm d}k\ \right |^2\ ,
\end{equation}
which is the analogue of equation (11) in \cite{Smi17}. The unitarity of $S$
guarantees the conservation of probability.
$$\sum _{h} \int_{-\infty}^{\infty} P_{h,h'}(s) {\rm d}s =1\ .  $$

For a Gaussian envelope,
\begin{equation}\label{gauss}
\omega(k) = \left(\frac{2}{\pi\sigma^2}\right)^{\frac{1}{4}}
e^{-\frac{(k-k_0)^2}{\sigma^2}}
\end{equation}
and under the condition $k_{0}>2\sigma$, one can approximate the
delay-time distribution by (see (17) in \cite{Smi17})
\begin{eqnarray}
\hspace{-14mm} P_{h,h'}(s)&\approx &\frac{1}{2\pi}  \int_{-\infty}^{\infty}{\rm d}\eta\ e^{-i \eta  s}
e^{-\frac{ \eta^2}{2 \sigma^2}} \times
  \nonumber  \\
& &\left \{ \sqrt{ \frac{2}{\pi\sigma^2}} \int_{-\infty}^{\infty}{\rm d}\xi\ e^{-\frac{2(\xi-k_0)^2}{\sigma^2}} S_{h,h'}\left(\xi+\frac{\eta}{2}\right)\ \overline{S_{h,h'}}\left(\xi-\frac{\eta}{2}\right)\right \}  .
\label{poftem2prime}
\end{eqnarray}
Using Eq.~(\ref{sfrompaths}) one can write an explicit expression
for $P_{h,h'}(s)$ for any values of $\sigma$ and $k_0$ which satisfy the
conditions underlying (\ref{poftem2prime}),
\begin{eqnarray}\label{dsumOrbs}
P_{h,h'}(s)&=&\sum_{n=0}^{\infty}\sum_{\alpha,\beta\in \mathcal{A}^{(n)}_{h,h'}} \left[e^{ik_0(l_{\alpha}-l_{\beta})}
e^{- (l_{\alpha}-l_{\beta})^2\sigma^2/8}\right ] A^{(n)}_{\alpha} \overline{A^{(n)}_{\beta}} \times \nonumber \\ & & \left [  \frac{\sigma}{\sqrt{2\pi}}   e^{-\left(\frac{l_{\alpha}+ l_{\beta}}{2}-s\right)^2\sigma^2/2}\right ].
\end{eqnarray}
In the above result, we did not include the reflections from the vertex $h'$
(which correspond to zero delay).  To render the discussion more transparent,
we shall proceed in the limit where $\sigma$ is very large, which allows to
write the first square bracket above as a Kronecker $\delta$ and the last
square bracket as a Dirac $\delta$ functions, resulting in
\begin{equation}\label{sumFam}
P_{h,h'}(s)=\sum_{n=0}^{\infty}\sum_{\alpha,\beta\in \mathcal{A}^{(n)}_{h,h'}}\delta_{l_{\alpha},l_{\beta}}  A^{(n)}_{\alpha} \overline{A^{(n)}_{\beta}}   \delta(l_{\alpha}-s)\ .
\end{equation}
\begin{figure}[b]
\centerline{\includegraphics[width=1.0\textwidth]{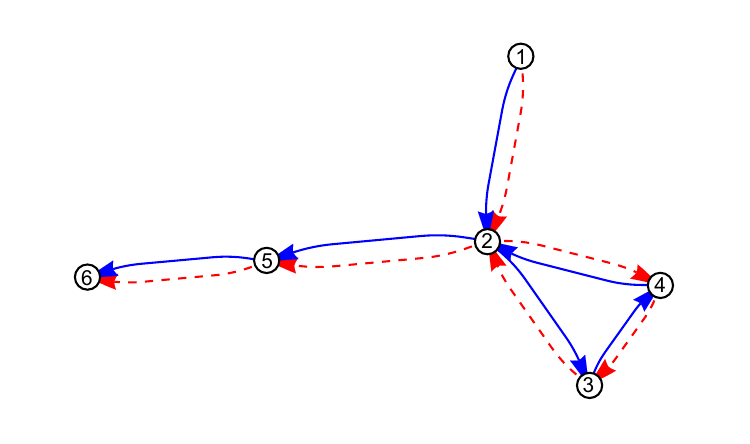}}
\caption{Isometric but topologically distinct paths differing in the
  orientation in which a loop of three vertices is traversed.}
\label{fig1}
\end{figure}
This expression can be simplified by recalling that the length of any path
$\alpha \in \mathcal{A}_{h,h'}^{(n)}$ can be written as $l_{\alpha}
=\sum_{e\in \mathcal{E}} q_e(\alpha) L_e$ where $q_e$ are non negative
integers whose sum is $n+1$. Note that $l_{\alpha}$ does not depend on the
direction in which the edges are traversed.  Because of the rational
independence of the edge lengths, paths which share the same length must share
also the same sequences $\{q_e\}_{e\in \mathcal{E}}$, and they are distinct if
they cross the same edges the same number of times but in different
order. Figure \ref{fig1} shows an example for such isometric but topologically
distinct paths. Denote ${\bf q}^{(n)}=\{q_e\}_{e\in \mathcal{E}}$ with $\sum
q_{e\in \mathcal{E}}=n+1$. The set of isometric paths which share the code
${\bf q}^{(n)}$ will be denoted by $\Gamma_{h,h'} ({\bf q}^{(n)})$. Then
\begin{eqnarray}\label{quantprob1}
P_{h,h'}(s)&=&\sum_{n=0}^{\infty}\sum_{{\bf q}^{(n)}} \left |\sum_{\gamma\in \Gamma_{h,h'} ({\bf q}^{(n)})}  A^{(n)}_{\gamma}\right |^2   \delta(l_{{\bf q}^{(n)}}-s)\\
&=&\sum_{n=0}^{\infty}\sum_{{\bf q}^{(n)}} p_{{\bf q}^{(n)}}   \delta(l_{{\bf q}^{(n)}}-s)\ .
\label{quantprob}
\end{eqnarray}
where the probabilities $p_{{\bf q}^{(n)}}$ contain all interference effects
between the isometric paths belonging to $\Gamma_{h,h'} ({\bf q}^{(n)})$. The
result of these interferences is determined by the phases of the individual
amplitudes $A^{(n)}_{\gamma}$. These in turn depend on the phases of the
elements of the vertex-scattering matrices $\sigma^{(v)}_{d_{j+1},d_j}$
encountered along the path $\gamma$ but they are independent of the precise
values of the edge lengths. Thus, the only information about the actual
lengths of the graph edges in the delay time distribution comes from the Dirac
delta functions concentrating at the path lengths $l_{{\bf
    q}^{(n)}}=\sum_{e\in\mathcal{E}}q_{e}L_{e}$.

It is convenient to define the cumulative probability
\begin{equation}
C_{h,h'}(s)= \int_{0}^{s}P_{h,h'}(t){\rm d}t = \sum_{n=0}^{\infty}\sum_{{\bf q}^{(n)}} p_{{\bf q}^{(n)}}   \Theta(s-l_{{\bf q}^{(n)}})\ ,
\label{commdist}
\end{equation}
where $\Theta (x)$ is the Heavyside function. Clearly, $C_{h,h'}(s)$ is a
non-decreasing function of s. On the other hand it depends parametrically on
the edge lengths $\mathcal{L}$ and is a non-increasing function of any
$L_{e}\ (e\in\mathcal{E})$, because these lengths appear only in the arguments
of the Heavyside step functions.  We can use this fact to bound $C_{h,h'}(s)$
from below and above by similar expressions with modified edge lengths. To
this end define the function
\begin{equation}\label{Cell}
C_{h,h'}(s,\ell)=
\sum_{n=0}^{\infty}\sum_{{\bf q}^{(n)}} p_{{\bf q}^{(n)}}
\Theta(s-(n+1)\ell)\,,
\end{equation}
where all edge lengths have been replaced by one and the same value
$\ell$. Note that this is a formal definition and not related to the delay
distribution of a graph with equal edge lengths, because Eq.~(\ref{commdist})
was derived under the assumption of rationally independent lengths. Take
now $\ell=\max (L_e)=\overline{L}$ being the maximum of the edge lengths of
the graph under consideration. For the same value of $s$ the arguments of the
Heavyside functions in Eq.~(\ref{Cell}) are smaller (or equal) in comparison
to Eq.~(\ref{commdist}) and thus in Eq.~(\ref{Cell}) less terms
contribute. Repeating this argument for $\min (L_e) = \underline{L}$ we see
that the cumulative delay distribution can be bound from below and above by
\begin{equation}
  C_{h,h'}(s;\overline{{L}})\le  C_{h,h'}(s)\le C_{h,h'}(s;\underline{{L}})\ .
\label{bounds}
\end{equation}
Note that $C_{h,h'}(s;\ell)=C_{h,h'}(s/\ell;1)$. Thus it suffices to calculate
$C_{h,h'}(t;1)$ for integer values of $t$. We refer to this quantity as the
cumulative probability for the {\em topological delay time} $t$, i.e. the
number of edges along the walk. Since there is no metric information to
consider, $C_{h,h'}(t;1)$ is typically easier to calculate than the full
expression in Eq.~(\ref{commdist}).

To proceed, write $P_{h,h'}(s)=P^{(D)}_{h,h'}(s)+P^{(ND)}_{h,h'}(s)$, where
\begin{eqnarray}
P^{(D)}_{h,h'}(s)=\sum_{n=0}^{\infty}\sum_{{\bf q}_n} \left [\sum_{\gamma\in \Gamma_{h,h'} ({\bf q}^{(n)})}  |A^{(n)}_{\gamma} |^2 \right ] \delta(l_{{\bf q}^{(n)}}-s)
\label{diagonal1}
\end{eqnarray}
\begin{eqnarray}
\label{nondiagonal}
P^{(ND)}_{h,h'}(s) =  \sum_{n=0}^{\infty}\sum_{{\bf q}_n} \left [
\sum_{\gamma\ne\gamma'\in \Gamma_{h,h'} ({\bf q}^{(n)})}  A^{(n)}_{\gamma}
 \overline{A^{(n)}_{\gamma'}}\right ]
  \delta(l_{{\bf q}^{(n)}}-s)\ .
\end{eqnarray}
The partition of the delay-time distribution into the Diagonal part
$P^{(D)}_{h,h'}(s)$ and the Non-Diagonal part $P^{(ND)}_{h,h'}(s)$ separates
the purely "classical" contribution from the contribution from the
interference of waves which propagate on isometric paths. The former will be
studied in the next section. Sometimes, (when e.g., $h\ne h'$ and the graph
is not invariant under geometrical symmetries) the contribution of
$P^{(ND)}_{h,h'}(s)$ can be ignored upon further averaging. However this is
not always the case, especially since the number of isometric trajectories $|
\Gamma_{h,h'} ({\bf q}^{(n)})|$ may increase indefinitely with $n$
\cite{Holger,SS00c,UriGavish}, and the sums do not necessarily vanish in spite of
the fact that the individual contributions have complicated, seemingly random
phases.

While some general properties of the classical time-delay distribution (\ref
{diagonal1}) can be derived as presented in the next section, there are no
analogous results pertaining to the complete expression in Eq.~(\ref
{quantprob}). However, in section \ref {sec:tjunction} we shall apply all
results of the present and the following section to a simple graph and derive 
analytical results for both, the classical and the quantum delay distribution.

\section{The classical delay-time distribution}
\label{diagonal}
In the present section we provide a classical description of the
  delay-time distribution. It is a valid approximation when quantum
  interference effects are negligible, either because of decoherence
  mechanisms in the scattering process or for short times, when the
  contributing trajectories do not have isometric partners. For long times and
  coherent dynamics a comparison to the reference provided by the classical
  description can highlight the features of the delay distribution which are
  due to genuine quantum (wave) properties of the scattering process, e.g.  an
  enhancement of long delay times (algebraic vs. exponential decay) in Section
  \ref{sec:tjunction}.

In the classical analogue of the scattering process described above,
one considers a classical particle which moves with a constant speed on the
incoming lead $h'$, and its probability to enter the graph through an edge
$d_0$ is $|\tau_{d_0,h'}|^2$. Reaching the next vertex after traversing a
distance $L_{d_0}$, it scatters into any of the connected edges $d_1$ with
probability $\tilde M_{d_1,d_0}$ (\ref{clasprob}) and so on until it leaves
the graph from the edge $d_n$ to the lead $h$ after being scattered on $n$
intermediate vertices. The length of the traversed trajectory between the
entrance and exit vertices is $l_{d_0,\cdots,d_n} =\sum_{j=0}^n
L_{{d_j}}$. Thus, the delay-time distribution is
\begin{equation}
\hspace{-15mm}
P_{h,h'}\cl(s) = \sum_{n=0}^{\infty}\sum_{d_{0},\cdots,d_{n}\in \mathcal {D}}
|\tau_{h,d_n}|^2\left\{\prod_{i=1}^n\tilde M_{d_i,d_{i-1}} \right\} | \tau_{d_0,h'}|^2\delta(s-l_{d_0,\cdots,d_n})
\ .
\label{dclass}
\end{equation}
This expression could be further reduced by grouping together
trajectories which share the same lengths, and the result reproduces the
expression for $P_{h,h'}^{(D)}(s)$ given in Eq.~(\ref{diagonal1}).

Again, it is convenient to define the cumulative probability,
\begin{align}
C_{h,h'}\cl(s) &= \int_0^{t}P_{h,h'}\cl(t)\,{\rm d}t \nonumber \\
&=\sum_{n=0}^{\infty}\sum_{d_{0},\cdots,d_{n}\in \mathcal {D}}
|\tau_{h,d_n}|^2\left\{\prod_{i=1}^n\tilde M_{d_i,d_{i-1}}\right\} |\tau_{d_0,h'}|^2 \ \Theta(s-l_{d_0,\cdots,d_n})
\ ,
\label{cclass}
\end{align}
in complete analogy to Eq.~(\ref {commdist}). Again the cumulative probability
is monotonically decreasing as a function of the edge lengths since all the
factors multiplying the Heavyside function in (\ref {cclass}) are
positive. Hence one can bound $C_{h,h'}\cl(s)$ in a similar way as in
(\ref{bounds}).

We will now derive the leading asymptotic behavior of
  $P_{h,h'}\cl(s)$ for large time. To this end we consider the
  Laplace transform of Eq.~(\ref{dclass})
\begin{align}
  \mathfrak{L}P_{h,h'}\cl(z)&=\int_{0}^{\infty}P_{h,h'}\cl(s)\,e^{-sz}
  \mathrm{d}s
  \\
  &=\sum_{d ,d'}
|\tau_{h,d}|^2 \sum_{n=0}^{\infty}(\tilde M^n(z))_{d,d'}\,e^{-zL_{d'}} |\tau_{d',h'}|^2 \nonumber \\
&=\sum_{d ,d'}  |\tau_{h,d}|^2 \left [I-\tilde M(z) \right]^{-1}_{d,d'}e^{-zL_{d'}} |\tau_{d',h'}|^2 \ ,
\label{dclass1}
\end{align}
where
\begin{equation}\label{Mz}
\tilde M(z)= e^{-zL}\tilde M \ ,
\end{equation}
and $L$ is a diagonal matrix with entries $L_d$. Note that according to
Eq.~(\ref{dclass1}) the poles of $\mathfrak{L}P_{h,h'}\cl(z)$ are related
to the zeroes of $\det(I-\tilde M(z))$ and the residues at these poles
can be computed with Jacobi's formula (adj = adjugate):
\begin{align}
\frac{\rm d}{{\rm d} z} \det \left (I^{(D)}-\tilde M(z) \right) &=
-{\rm {tr}} \left [ {\rm {adj}}\left (I^{(D)}-\tilde M(z) \right)
\frac{{\rm d}}{{\rm d} z}\tilde M(z)\right ]
 \\&=
   {\rm {tr}} \left [ {\rm {adj}}\left (I^{(D)}-\tilde M(z) \right) L
     \,{\tilde M(z)}
     \right ]\,.
   \label{resi}
\end{align}
The idea is now to use this information about the the analytic properties of
$\mathfrak{L}P_{h,h'}\cl(z)$ in order to invert the Laplace transform
by a complex contour integral. This procedure can be put on a solid basis by
applying the Wiener-Ikehara theorem to Eq.~(\ref{dclass1}). Using the results
of \cite{avner} one gets
\begin{align}
P_{h,h'}\cl(s) \approx  e^{-s\xi}\  \sum_{ d,d' } |\tau_{h,d }|^2 \frac{\left [ {\rm   adj} \left ( I^{(D)}-\tilde M(-\xi) \right )\right ]_{d ,d'} e^{\xi L_{d'}}}
{ {\rm tr} \left [{\rm adj} \left (I^{(D)}-\tilde M(-\xi)\right ) L
{\tilde M(-\xi)}
    \right ] }
  |\tau_{d',h'}|^2 \quad(s\to\infty)
 \label{finalresult}
\end{align}
where $\xi$ is the largest real zero of ${\rm {det}}\left (I^{(D)}-\tilde M(z)
\right)$. It depends on both the graph connectivity and the set of edge
lengths $L$.  Eq.~(\ref{finalresult}) is the main result of the present
section. 

\section{Example}\label{sec:tjunction}
\def\tj{{\sf T}-junction}
\subsection{The \tj\ model.}
\begin{figure}[htb]
 \centerline{\includegraphics[width=6cm]{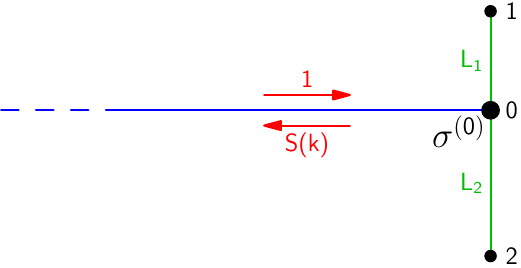}}
 \caption{\label{fig:Tj} A simple model graph consisting of two edges (green)
   and one scattering lead attached to the central vertex 0 (blue). In
   numerical calculations we use $L_{1}=(1+\sqrt{5})/8\approx0.4045$ and
   $L_{2}=1-L_{1}\approx0.5955$.}
\end{figure}

As an example we choose a graph which is simple enough to allow for an
analytical treatment and still rich enough to exhibit all aspects of the
theory outlined above. In particular the model demonstrates the influence of
quantum interferences on the delay distribution, $P^{(ND)}(s)$ from
Eq.~(\ref{nondiagonal}). The graph consists of two edges ($E=2$, $D=4$) which
are connected at a central vertex. Moreover, at this vertex a single
scattering lead is attached. Thus the central vertex has the total degree
three. Both internal edges end in vertices of degree one with Neumann b.c.
The graph can be depicted as shown in Fig.~\ref{fig:Tj} and we refer to it as
a \tj. In order to specify the model completely we need to define the lengths
of the two edges and the 3$\times$3 scattering matrix of the central
vertex. For the lengths we choose two rationally independent values such that
the total length is $L=L_{1}+L_{2}=1$. This is no restriction of generality as
the delay time scales proportionally to this quantity. Our choice for
$\sigma^{(0)}$ is motivated by analytical simplicity,
\begin{eqnarray}\label{sigma0}
  \sigma^{(0)}&=&\frac{1}{2}\begin{pmatrix}
    0 & +\sqrt{2} & -\sqrt{2} \\
    -\sqrt{2} & 1 & 1 \\
    +\sqrt{2} & 1 & 1 
\end{pmatrix}\,.
\end{eqnarray}
Here the lower right 2$\times$2 block describes the scattering within the
interior of the graph. Our calculations are simplified by the fact that in
this block no phases must be considered. The first column and the first row
contain the transition amplitudes $\tau$ from the scattering lead into the
graph and back. The amplitude at the central vertex for a direct back
scattering into the lead is zero, $\rho_{00}=\sigma^{(0)}_{0,0}=0$.

Note that according to \cite{KS00a} any choice of a unitary scattering matrix
$\sigma^{(0)}(k_{0})$ at some {\em fixed} wave number $k_{0}$ is compatible
with a self-adjoint Laplacian. However, this choice also fixes the variation
of $\sigma^{(0)}(k)$ with wave number which depends on the parameter
$(k-k_{0})/(k+k_{0})$ \cite{KS00a}. As we consider here an envelope function
with a width $\sigma\ll k_{0}$ we can approximate
$\sigma^{(0)}(k)\approx\sigma^{(0)}(k_{0})$ and ignore the energy dependence
of the vertex scattering matrix.

\subsection{The S-matrix.}
Using Eq.~(\ref{defS}) we can now derive an expression for $S(k)$. The indices
$h,h'$ from Eq.~(\ref{defS}) can be omitted, since there is just a single
scattering channel. Defining $\phi_{1,2}(k)=e^{2ikL_{1,2}}$ we obtain
\begin{align}\label{stj}
  S(k)&=
  \frac{\phi_{1}\phi_{2}-\frac{\phi_{1}+\phi_{2}}{2}}
       {1-\frac{\phi_{1}+\phi_{2}}{2}}
\\&=\sum_{t_{1},t_{2}=1}^{\infty}
\frac{(t_{1}+t_{2})-(t_{1}-t_{2})^{2}}{2^{t_{1}+t_{2}}\,t_{1}t_{2}}
\binom{t_{1}+t_{2}-2}{t_{1}-1}\,\phi_{1}^{t_{1}}\phi_{2}^{t_{2}}
-\sum_{t=1}^{\infty}\frac{\phi_{1}^{t}+\phi_{2}^{t}}{2^{t}}
\label{stj2}
\end{align}
(see \ref{app:quantum} for details). The first line of is a compact
representation which is suitable for numerical calculations and clearly
highlights the resonance structure of the scattering matrix. The second line
is an expansion of $S(k)$ in terms of families of isometric trajectories
starting and ending on the scattering lead. These families are labelled by
pairs $\alpha=(t_{1},t_{2})$ counting the number of reflections from the the
first and second outer vertex, respectively. Trajectories which are restricted
to a single edge are accounted for by the second sum. In the notation of
Eq.~(\ref{quantprob1}) the numbers {\bf q} defining a family count the
traversals of directed bonds. However, in our simple model, an edge is always
traversed outward and inward successively, thus $q_{0\to1}=q_{1\to0}=t_{1}$
and $q_{0\to2}=q_{2\to0}=t_{2}$. We will refer to the integer value
$t=t_{1}+t_{2}$ as the topological time of a path on the \tj\ graph.  As in
Eq.~(\ref{sfrompaths}) the oscillating phase factors
$\phi_{1}^{t_{1}}\phi_{2}^{t_{2}}=\exp(ikl_{t_{1},t_{2}})$ in Eq.~(\ref{stj2})
depend on the total length of the trajectories within a family,
\begin{equation}\label{ltr}
  l_{t_{1},t_{2}}=2(t_{1}L_{1}+t_{2}L_{2})\,,
\end{equation}
while the rational prefactors represent the sum of amplitudes from all
trajectories within a family, as in Eq.~(\ref{quantprob1}).

\begin{figure}[ht]
\centerline{\includegraphics[width=1.0\textwidth]{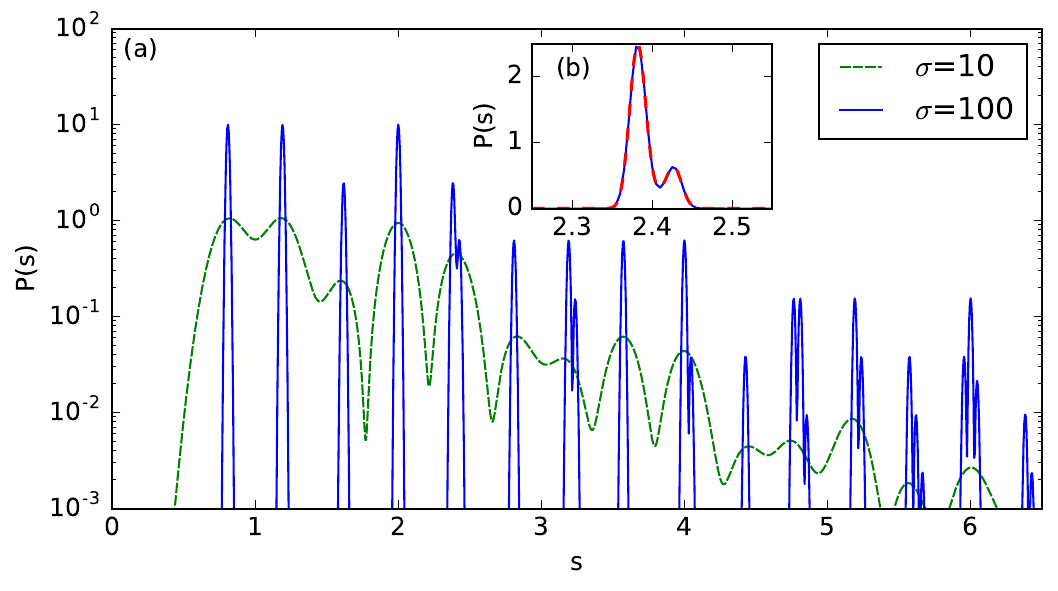}}
\caption{(a) The probability density $P(s)$ for a \tj\ with
  $L_{1}=(1+\sqrt{5})/8\approx0.4045$ and $L_{2}=1-L_{1}\approx0.5955$.  is
  shown for $\sigma=100$ (blue) and $\sigma=10$ (green broken line) on a
  logarithmic scale. (b) The inset enlarges a region where two peaks with
  almost degenerate trajectory lengths interfere ($\sigma=100$). The dashed
  red line in the inset is the interference pattern predicted by
  Eq.~(\ref{dsumOrbs}).}
\label{fig:interf}
\end{figure}

Fig.~\ref{fig:interf} shows the time delay density computed with
Eqs.~(\ref{ps}), (\ref{gauss}), (\ref{stj}) by a Fourier transform of the
scattering matrix $S(k)$. The two curves correspond to two different envelope
widths $\sigma$. As predicted above in Eqs.~(\ref{dsumOrbs}), (\ref{sumFam}) a
series of sharp peaks centered at the lengths of scattering trajectories
develops as $\sigma$ grows. For example, the first two peaks at
$s=2L_{1}\approx0.81$ and $s=2L_{2}\approx1.19$ each correspond to a single
scattering trajectory which enters the graph, visits one of the outer vertices
1 or 2 and returns to the lead. However, to most of the peaks more than one
trajectory contributes and their interference, expressed by the rational
prefactors in Eq.~(\ref{stj2}), determines the height of the peak. For growing
time $s$, an increasing fraction of peaks have a separation of the order of
$\sim\sigma^{-1}$ or smaller and overlap. This is a limitation to
Eq.~(\ref{sumFam}) and the subsequent theory. An example at $3L_{1}\approx
2L_{2}\approx 2.4$ is magnified and compared to the prediction of
Eq.~(\ref{dsumOrbs}) in the inset Fig.~\ref{fig:interf}(b).

\subsection{The topological delay time distribution.}\label{sec:tjtopdel}
Within the asymptotic approximation for broad envelope functions (short
pulses), Eq.~(\ref{sumFam}), we can evaluate the (cumulative) distribution
of delay times (\ref{commdist}) for the \tj. According to
Eqs.~(\ref{quantprob1})-(\ref{commdist}) the squared coefficients from 
Eq.~(\ref{stj2}) provide the weigth of a family and we obtain
\begin{align}\nonumber
C(s)=&
\sum_{t_{1},t_{2}=1}^{\infty}
\left(\frac{(t_{1}+t_{2})-(t_{1}-t_{2})^{2}}{2^{t_{1}+t_{2}}\,t_{1}t_{2}}
\binom{t_{1}+t_{2}-2}{t_{1}-1}\right)^{2}\,\Theta(s-l_{t_{1},t_{2}})
\\&\qquad+\sum_{t=1}^{\infty}2^{-2t}\,[\Theta(s-l_{t,0})+\Theta(s-l_{0,t})]\,.
\label{Cs}
\end{align}
As in Eq.~(\ref{bounds}), this function can be bound from below and above by a
variation of the edge lengths. Define $C(s,\ell)$ to denote the r.h.s of
Eq.~(\ref{Cs}) with both edge lengths $L_{1}$, $L_{2}$ replaced by some value
$\ell$ such that $l_{t_{1},t_{2}}$ is $2(t_{1}+t_{2})\ell$. Then the Heavyside
functions in Eq.~(\ref{Cs}) are $\Theta(s-2t\ell)$ and select all terms with
topological times $t=t_{1}+t_{2}$ up to $\lfloor{s/2\ell}\rfloor$ (the largest
integer below $s/2\ell$). Thus, if $p_{t}$ denotes the sum of coefficients of
all terms with some fixed topological time $t$, $C(s,\ell)$ is the cumulant
sum
\begin{equation}
C(s,\ell)=\sum_{t=0}^{\lfloor{s/2\ell}\rfloor}p_{t}
\end{equation} 
Starting with the substitution $t_{2}=t-t_{1}$ we can evaluate $p_{t}$ as
\begin{align}\label{pt1}
  p_{t}&=2^{1-2t}+\sum_{t_{1}=1}^{t-1}\left(2^{-t}\frac{t-(t-2t_{1})^{2}}{t_{1}(t-t_{1})}
  \binom{t-2}{t_{1}-1}\right)^{2}
\\&=\frac{3}{4}\frac{4^{2-t}}{t(t-1)}\binom{2t-4}{t-2}
\qquad(t>1)\label{pt2}
\\&\approx\frac{3}{4}\frac{t^{-5/2}}{\sqrt{\pi}}\hspace*{24mm}(t\to\infty)
\label{pt3}
\end{align}
while $p_{0}=0$ and $p_{1}=1/2$. Eq.~(\ref{pt2}) can be found with the help of
standard computer algebra, and a formal proof can be based
  on the methods outlined in \cite{A=B}. $p_{t}$ is a normalized discrete
probability distribution (the distribution of topological time delays) and
its cumulant sum is
\begin{align}\label{ct}
  c_{t}&=\sum_{t'=0}^{t}p_{t'}
  \\&=1-\frac{2}{4^{t}t}\binom{2t-2}{t-1}
  \\&\approx 1-t^{-3/2}/\sqrt{4\pi}\qquad(t\to\infty)\,.
  \label{ctasymp}
\end{align}
Now consider $C(s,L_{1})$ and $C(s,L_{2})$. Assuming without loss of
generality $L_{1}<L_{2}$ we have $2(t_{1}+t_{2})L_{1}\le
l_{t_{1},t_{2}}\le2(t_{1}+t_{2})L_{2}$, i.e. in comparison with $C(s)$ the
Heaviside steps occur in $C(s,L_{1})$ for smaller and in $C(s,L_{2})$ for
larger values of $s$ while the coefficients remain unchanged. Hence
\begin{equation}\label{tjbounds}
C(s,L_{2})\le C(s)\le C(s,L_{1})
\end{equation}
Asymptotically for large delay $s\to\infty$ these bounds on $C(s)$ are
explicitly given by substitution of $s/2L_{1,2}$ into Eq.~(\ref{ctasymp}),   
\begin{equation}\label{tjboundsexpl}
  1-\frac{(s/2L_{2})^{-3/2}}{\sqrt{4\pi}}\le C(s)\le1-\frac{(s/2L_{1})^{-3/2}}{\sqrt{4\pi}}\qquad(s\to\infty)\,.
\end{equation}
We conclude that the probability $1-C(s)$ to measure a delay larger than $s$
falls off as a power law with exponent $-3/2$ and that for
$2L_{1}\lesssim1\lesssim2L_{2}$ a prefactor $1/\sqrt{4\pi}$ should be expected. 

\begin{figure}[htb]
\centerline{\includegraphics[width=1.0\textwidth]{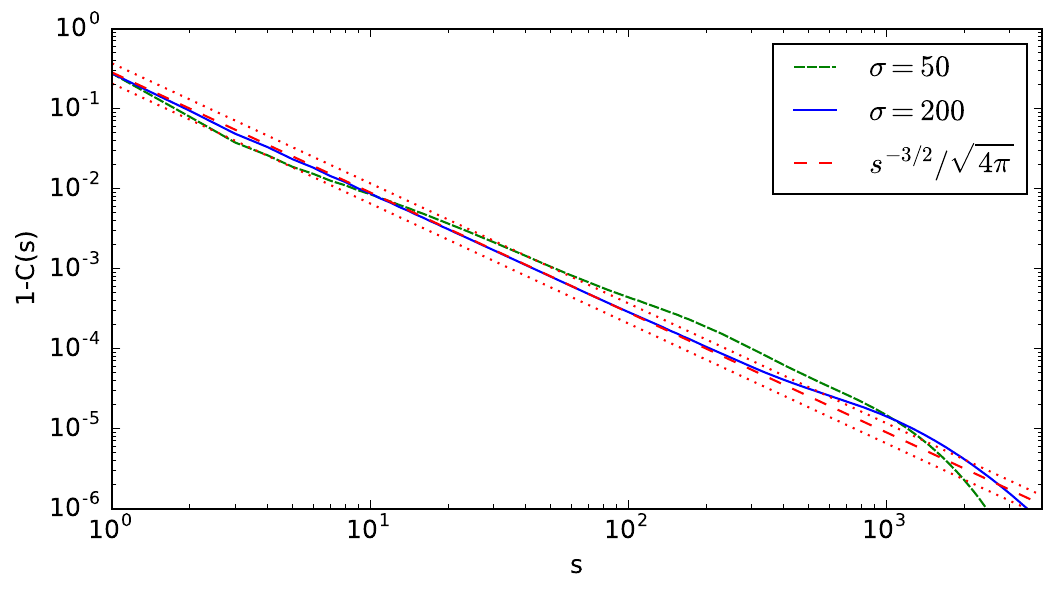}}
\caption{The long-time tail of the integrated delay-time distribution is shown
  on a double logarithmic scale for $\sigma=50$ (broken green line) and
  $\sigma=200$ (blue). The dashed red line is the theoretical result
  (\ref{csreso}) based upon the distribution of narrow resonances. The upper
  and the lower dotted lines are the bounds (\ref{tjboundsexpl}) derived from
  the topological delay time distribution. For the calculations we used
  $k_{0}=1.000$ and a discrete Fourier transform of the S-matrix (\ref{stj})
  on a grid with a spacing $\delta k\sim10^{-4}$ which ensures convergence of
  the distribution in the displayed region $s\le4.000$.  }
\label{fig:tjQlt}
\end{figure}

\subsection{The long-time delay distribution.}
For $s\to\infty$ the factor $e^{-iks}$ in the Fourier integral of
Eq.~(\ref{ps}) has very fast oscillations which cancel out unless $S(k)$ is
rapidly changing too. Therefore the asymptotic time delay for large $s$ is
related to narrow resonances of the scattering matrix. On this basis we can
develop an alternative approach to the delay time distribution, similar to
\cite{DHM92,HAO09}. In \ref{app:reso} we show that $C(s)$ for
large $s$ can be approximated by the sum
\begin{align}\label{csresosum}
C(s)&=1-4\pi\sum_{n}\omega^{2}(\kappa_{n})\gamma_{n}e^{-2\gamma_{n}s}\,,
\end{align}
where $\kappa_n-i\gamma_{n}$ are the poles of the scattering matrix
(\ref{stj}) in the complex $k$-plane. For broad envelope functions $\omega(k)$
many resonances contribute and we can approximate Eq.~(\ref{csresosum}) by an
integral over the resonant wave number $\kappa$ and the resonance width
$\gamma$,
\begin{align}\label{csresoint}
  C(s)&=1-4\pi\int_{0}^{\infty}d\kappa\int_{0}^{\infty}d\gamma\,\rho(\kappa,\gamma)\,
  \omega^{2}(\kappa)\gamma\,e^{-2\gamma s}
  \\&=1-\frac{1}{\sqrt{4\pi}}\left(\frac{s}{L}\right)^{-3/2}\,,\label{csreso}
\end{align}
where $L=L_{1}+L_{2}$ is the total length of the graph,
\begin{equation}\label{resodens}
  \rho(\kappa,\gamma)=\frac{1}{\pi^{2}}\sqrt{\frac{L^{3}}{2\gamma}}
\end{equation}
is the average density of resonances in the complex plane and the
normalization of $\omega(k)$ was used to integrate over $\kappa$ (see
\ref{app:reso} for details). Clearly, Eq.~(\ref{csreso}) is compatible with
Eq.~(\ref{tjboundsexpl}) and even refines this prediction from the previous
subsection. Moreover it becomes clear, that a condition for this result is
that the envelope function covers many resonances with a relevant
contribution in Eq.~(\ref{csresosum}), i.e. with a width up to $\gamma(s)\sim
s^{-1}$. Since $\rho(k,\gamma)\sim \gamma^{-1/2}$ the number of contributing
resonances scales as $\sigma\sqrt{\gamma(s)}$ and we infer that
Eq.~(\ref{csresoint}) is valid up to a maximum time $s\sim\sigma^{2}$. Beyond
that value $C(s)$ will have a non-universal behaviour dictated by the
resonances with the smallest widths which are covered by the envelope
function.

Fig.~\ref{fig:tjQlt} illustrates the results from the previous and the present
subsections. In order to highlight the power-law tail of the delay time
distribution we show the quantity $1-C(s)=\int_{s}^{\infty}ds' P(s')$,
i.e. the probability to measure a delay exceeding $s$. We compare numerical
results for $\sigma=50$ and $\sigma=200$ to the bounds derived from the
topological delay time in Section \ref{sec:tjtopdel} and to Eq.~(\ref{csreso})
above. For $\sigma=200$ there is a very good agreement up to
$s\sim300$. Beyond $s=3.000$ $P(s)$ falls off
very fast because the region covered by the envelope function contains no
resonances which are narrow enough to contribute. For smaller $\sigma$ the
deviations set in earlier and are generally larger, as expected.  

\begin{figure}[htb]
\centerline{\includegraphics[width=1.0\textwidth]{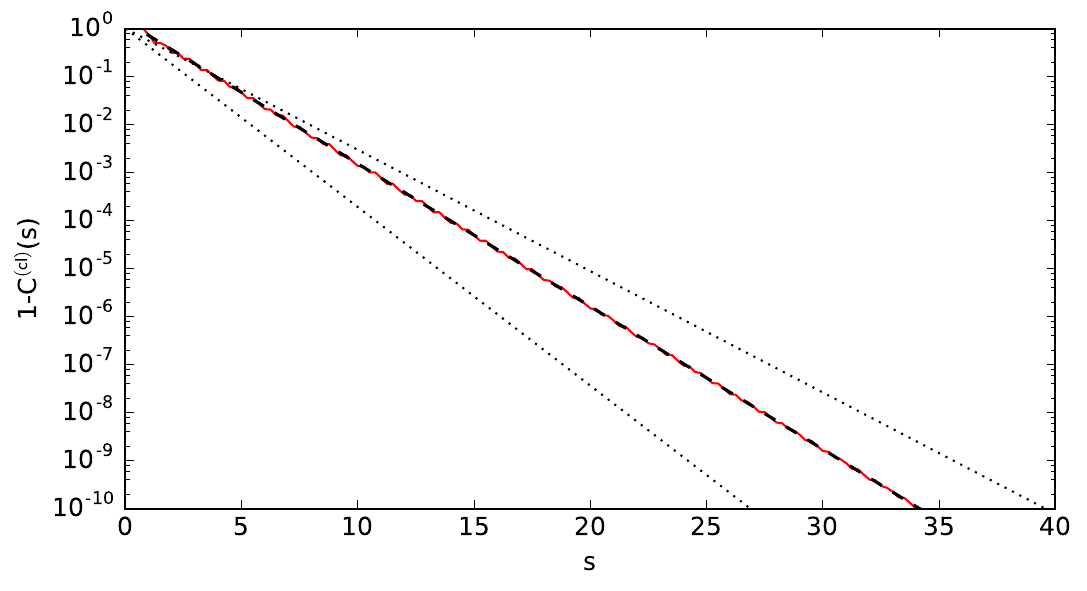}}
\caption{The classical delay-time distribution is shown on a logarithmic scale
  with a red line. The dashed line is the theoretical result (\ref{cltj}) for
  $\xi\approx0.6846$ (the numerical root of Eq.~(\ref{tjdetxi})). The dotted
  lines are the upper and the lower bounds on $\ccl(s)$ derived from the
  topological delay distribution.  }
\label{fig:cltj}
\end{figure}

\subsection{The clasical delay distribution.}\label{sec:cltj}
According to Eq.~(\ref{cclass}), for the clasical delay distribution we have
to sum over all paths leading from the scattering channel into the graph and
back to the channel. For the \tj\ these paths consist of $t_{1}$ excursions
from the central vertex 0 to vertex 1 and $t_{2}$ excursions to vertex 2, in
arbitrary order.  The total length of such a path was given in
Eq.~(\ref{ltr}). The product of matrix elements of $\tilde M$ along the path
is $4^{-(t-1)}$, corresponding to $t-1$ inner crossings of vertex 0 (see
\ref{app:classical} for details). Again $t=t_{1}+t_{2}$ denotes the
topological time. Together with the probabilities
$|\tau_{h,d_{0}}|^{2}=|\tau_{h,d_{n}}|^{2}=1/2$ for entering and leaving the
interior graph from/to the scattering channel the weight of each path is
$4^{-t}$. The number of paths with given $t_{1}$ and $t_{2}$ is easily counted
and thus from Eq.~(\ref{cclass}) we find for the \tj
\begin{equation}\label{cclsum}
\ccl(s)=\sum_{t_{1},t_{2}}^{\infty}4^{-(t_{1}+t_{2})}\binom{t_{1}+t_{2}}{t_{1}}\Theta(s-l_{t_{1},t_{2}})\,.
\end{equation}
Similar to Eq.~(\ref{tjbounds}) we can estimate this quantity by substitution
of a common value $\ell$ for the edge lengths. As there are $2^{t}$ paths with
topological time $t$ we have
\begin{align}
  \ccl(s,\ell)&=\sum_{t=1}^{\infty}2^{-t}\Theta(s-2t\ell)
  \\&=1-2^{-\lfloor s/2\ell\rfloor}\,.
\end{align}
With $\ell=L_{1}$ ($\ell=L_{2}$) this expression is an upper (lower) bound for
$\ccl(s)$. However, a much more precise estimate can be obtained from
Eq.~(\ref{finalresult}). For the \tj\ we find 
\begin{align}\label{tjdetxi}
  \det(1-\tilde M(-\xi))=1-\frac{1}{4}e^{2L_{1}\xi}-\frac{1}{4}e^{2L_{2}\xi}=0
\end{align}
and can solve for $\xi$. Then, integrating Eq.~(\ref{finalresult})
with respect to $s$ we have
\begin{align}
  \ccl(s)&=1-\int_{s}^{\infty}ds'\,P\cl(s')
  \\&\approx1-\frac{A(\xi)}{\xi}e^{-\xi s}\qquad(s\to\infty)\label{cltj}
\end{align}
where  
\begin{align}\label{wxi}
  A(\xi)&=\frac{2}{L_{1}\,e^{2\xi L_{1}}+L_{2}\,e^{2\xi L_{2}}}
\end{align}
dentotes the sum in Eq.~(\ref{finalresult}) for a \tj. See 
\ref{app:classical} for more details on the derivation of these results.

Note that Eq.~(\ref{tjdetxi}) requires a numerical solution in general. A full
analytical soltion can be given, e.g., for a \tj\ with two edges of equal
length, $L_{1}=L_{2}=1/2$. Expanding around this trivial case to leading order
in the difference of the edge lengths $\delta L=L_{2}-L_{1}$ for fixed
$L_{1}+L_{2}=1$ one finds $\xi\approx\ln2\cdot(1-\frac{\ln2}{2}[\delta
  L]^{2})$ and $A(\xi)=1-\ln2\cdot[\delta L]^{2}$.

Fig.~\ref{fig:cltj} illustrates our results for the classical delay
distribution and shows a very accurate agreement between Eq.~(\ref{wxi}) and
numerical data generated from Eq.~(\ref{cclsum}).

\section{Conclusions} In the preceding sections we have provided a
  theory for the computation of the delay time distribution in scattering from
  quantum (wave-guide) networks. A main result was the reduction of the
  distribution to a purely combinatorial expression, the topological
  delay time distribution of Eq.~(\ref{Cell}). It provides bounds for the
  actual distribution which do not depend on the precise lengths of the
  edges of the network as long as they are not rationally related.
  
  In the last chapter we have given a complete solution for a simple
    graph, which reveals remarkable features. The coherent delay time
    distribution decays as a power-law while the classical distribution shows
    the expected exponential decay, emphasizing the importance of
    interferences effects when the scattering region supports a complex
    internal dynamics. From another perspective the algebraic decay is related
    to a particular distribution of the widths of long-lived scattering
    resonances which in this simple model was analytically accessible.

    The methods developed in the present paper and tested in the toy
      model of Section \ref{sec:tjunction} can now be applied to quantum
      graphs with a physically more interesting and challenging structure. To
      name an example, scattering from random non compact graphs is now under
      study, showing the effects of Anderson localization in the time
      domain. The results will be reported shortly.  

\section{Acknowledgements}
  Professor Steve Anlage is acknowledged for directing us to the subject, by
  sharing his preliminary experimental results. US acknowledges support from
  the Humboldt foundation.

\begin{appendix}  
\section{}\label{app:quantum}
Here we evaluate the scattering matrix for the \tj-model of Section
\ref{sec:tjunction} starting from Eq.~(\ref{defS}) and
Eq.~(\ref{sigma0}). There is only a single scattering channel $h=h'=0$. The
amplitude for direct reflection $$\rho_{0}=0$$ is given by the first element
of the matrix $\sigma^{(0)}$ in Eq.~(\ref{sigma0}).

The transition amplitudes $\tau_{d0}$ from the scattering channel into the
graph are non-zero only if the directed bond $d$ points outward from the
central vertex ($0\to1$, $0\to2$). Vice versa the transition amplitudes
$\tau_{0d}$ are non-zero for inward pointing bonds ($1\to0$, $2\to0$).
According to Eq.~(\ref{sigma0}) each
non-zero transition amplitude is $\pm\sqrt{2}/2$, and their product has
negative/positive sign if the first and the last edge traversed inside the
graph are equal/distinct.

The matrix $W$ has dimension 4\footnote{ Because of the bipartite structure of
  the graph with respect to inward/outward bonds it would be possible to
  reduce the whole calculation to 2$\times$2 matrices. We chose not to do so
  here in order to keep the notation parallel to the general result in
  Eq.~(\ref{defS}).} and is explicitly given by
\begin{equation}\label{Wtj}
W(k)=\begin{pmatrix}
    0&0&\frac{1}{2}e^{ikL_{1}}&\frac{1}{2}e^{ikL_{1}}\\[1mm]
    0&0&\frac{1}{2}e^{ikL_{2}}&\frac{1}{2}e^{ikL_{2}}\\[1mm]
    e^{ikL_{1}}&0&0&0\\[1mm]
    0&e^{ikL_{2}}&0&0 
\end{pmatrix}
\end{equation}
where we have ordered the four directed bonds of the graph such, that the
first two entries correspond to bonds from the central vertex outward and the
last two entries to bonds directed inward.  For compact notation we
define
\begin{equation}
  \phi_{1,2}(k)=e^{2ikL_{1,2}}
\end{equation}
and find $\det(I-W)=1-(\phi_{1}+\phi_{2})/2$. Now it is possible to calculate
$(I-W)^{-1}$ using the adjugate of $I-W$.
In fact, it suffices to calculate the
lower left 2$\times$2 block of the adjugate (outward to inward)
$$\frac{1}{2}\begin{pmatrix}
e^{ikL_1}(2-e^{2ikL_{2}})&e^{ik(2L_{1}+L_{2})}\\
e^{ik(L_{1}+2L_{2})}&e^{ikL_2}(2-e^{2ikL_{1}})\\
\end{pmatrix}$$
because only for this combination the product $\tau_{h,d}\tau_{d',h}$ in
Eq.~(\ref{defS}) is non-zero and equal to $\pm1/2$ (minus on the diagonal of
the block). According to Eq.~(\ref{defS}), the first and second column are
also multiplied by $e^{ikL_{1}}$ and $e^{ikL_{2}}$, respectively. Summation of
all four matrix elements finally yields Eq.~(\ref{stj}).  In order to arrive
at Eq.~(\ref{stj2}) we can expand the denominator as a geometric series and
regroup all terms according to the powers in $\phi_{1}$ and $\phi_2$.

As an alternative, Eq.~(\ref{stj2}) can also be obtained directly from a
summation of all paths on the graph as in Eq.~(\ref{sfrompaths}). A path
consists of several excursions from the central vertex 0 to either vertex 1 or
vertex 2 and back to zero. Each such excursion contributes a phase $\phi_{1}$
or $\phi_{2}$, respectively. Moreover, there is a transion amplitude $1/2$ for
every internal transition across vertex 0 and an amplitude $\pm1/\sqrt{2}$ for
a transition from the scattering channel into the graph and back. Therefore
each path with $t_{1}+t_{2}$ excursions has an amplitude
$\pm2^{-(t_{1}+t_{2})}$. Paths of the form 1\dots2 or 2\dots1 have a positive
sign and are counted by choosing the positions of the $n_{1}-1$ remaining
excursions to vertex 1 from the $t_{1}+t_{2}-2$ available inner time
steps. Paths of the form 1\dots1 or 2\dots2 have negative sign and are counted
in an analogous way.  We obtain
\begin{equation}\label{sfrompathstj2}
  S=\sum_{t_{1},t_{2}=0}^{\infty}
  \left[2\binom{t_{1}+t_{2}-2}{t_{1}-1}-\binom{t_{1}+t_{2}-2}{t_{1}-2}-\binom{t_{1}+t_{2}-2}{t_{2}-2}\right]\,\frac{\phi_{1}^{t_{1}}\phi_{2}^{t_{2}}}{2^{t_{1}+t_{2}}}\,.
\end{equation}
After applying binomial recursion (Pascal's triangle) to the second and the
third binomial, the first binomial can be factored out and the equivalence to
Eq.~(\ref{stj2}) is easily established.

\section{}\label{app:reso}
\def\k{{\kappa}}\def\dk{\delta k}

As obvious from Eq.~(\ref{stj}), the scattering matrix has a singularity if
$\phi_{1}(k)+\phi_{2}(k)=2$. For real $k>0$ this equation has no solution
since it would imply $\phi_{1}(k)=\phi_{2}(k)=1$, i.e. $kL_{1,2}=2m_{1,2}\pi$
and $L_{1}/L_{2}=m_{1}/m_{2}$ for integer $m_{1}$ and $m_{2}$. This is
excluded by the incommensurability of the bond lengths. However it is
possible that the two phases pass through a multiple of $2\pi$,
\begin{equation}\label{k1k2}
  e^{2ik_{1}L_{1}}=e^{2ik_{2}L_{2}}=1\,.
\end{equation}
at two different wave numbers $k_{1}$ and $k_{2}$ which have a very small
spacing
\begin{equation}
  \dk=|k_{1}-k_{2}|\,.
\end{equation}
We define
\begin{align}\label{lambda}
  L&=L_{1}+L_{2}\\
  \lambda&=\frac{L_{1}L_{2}}{L}
  \\\label{gamma}
\gamma&=\frac{\lambda^{2}}{L}\dk^{2}
\end{align}
and a weighted average of $k_{1}$ and $k_{2}$,  
\begin{align}\label{kappa}
  \k&=\frac{k_{1}L_{1}+k_{2}L_{2}}{L}
\end{align}
It is easy to verify that $S(k_{1})=S(k_{2})=-1$ and $S(\kappa)=+1$, i.e. the
phase of the S-matrix completes a full cycle in the small interval between
$k_{1}$ and $k_{2}$. In the immediate vicinity of $\kappa$ the functional form
of the phase is universal when $k-\kappa\sim\dk^{2}$. Namely, using
Eq.~(\ref{k1k2}) we have 
\begin{align}
  \phi_{1}(k)&=e^{2ik L_{1}}\approx 1+2i\lambda\dk+2iL_{1}\delta
  k-2\lambda^{2}\dk^{2}
  \\\phi_{2}(k)&=e^{2ik L_{2}}\approx 1-2i\lambda\dk+2iL_{2}\delta
  k-2\lambda^{2}\dk^{2}
  \\S(k)&
  \approx-\frac{k-(\kappa+i\gamma)}{k-(\kappa-i\gamma)}\label{sapp}
  \\&=e^{2i\arctan([k-\kappa]/\gamma)}\label{sapp1}\,.
\end{align}
Note that due to cancellations Eq.~(\ref{sapp}) is valid to leading order in
$\dk$ only, although $\phi_{1,2}(k)$ were expanded to second order. From this
result it is obvious that the scattering matrix has a pole close to the real
axis at $\kappa-i\gamma$. For a resonance of width $\gamma$ the maximal
derivative of the phase in Eq.~(\ref{sapp1}) is $2/\gamma$. Up to this value
of $s$ the Fourier integral in Eq.~(\ref{ps}) has a point of stationary phase
and thus a relevant contribution to $P(s)$ results. In the vicinity of
$\kappa$ we can approximate the envelope function by the constant
$\omega(\kappa)$. The resulting contribution is then found from the residue
$2i\gamma\,e^{-i\kappa s-\gamma s}$ of the remaining integrand at the
pole. Summation over all resonances gives
\begin{align}
  P(s)&=\frac{1}{2\pi}\left|4\pi\sum_{n}\gamma_{n}\omega(\kappa_{n})\,e^{-i\kappa_{n} s-\gamma_{n} s}\right|^{2}\,.
\end{align}
In this expression the contributions from different resonances $n$ to $P(s)$
will interfere. However, in $C(s)$ the integration with respect to $s$ will
destroy these interferences. To see this, expand $|\dots|^{2}$ as a double sum
over $n, n'$. Then nondiagonal terms have oscillating phase factors
$e^{i(\kappa_{n}-\kappa_{n'})s}$ and are suppressed in comparison to the
diagonal terms $n=n'$. We are left with
\begin{align}
  C(s)&=1-8\pi\int_{s}^{\infty}ds'\sum_{n}\omega^{2}(\kappa_{n})\gamma_{n}^{2}e^{-2\gamma_{n}s}
\end{align}
which finally yields Eq.~(\ref{csresosum}).  Further the sum over resonances
can be replaced by the integral Eq.~(\ref{csresoint}) if the envelope function
is broad and a large number of resonances contribute. In this way the delay
distribution for long times is related to the density of narrow resonances in
the complex plane. In order to estimate this density $\rho(\kappa,\gamma)$ we
first note that points $k_{1}$ with $e^{2ik_{1}L_{1}}$ have a density
$L_{1}/\pi$. At these points the second phase $\varphi_{2}(k_{1})=2k_{1}L_{2}$
can be treated as a random number with uniform distribution between $\pm\pi$.
If $|\varphi_{2}|$ is small, a small change $\dk=-\varphi_{2}/2L_{2}$ is
sufficient to bring it to zero. Thus a spacing between 0 and $\dk$ results
with probability $2\dk L_{2}/\pi$. Then $2L_{1}L_{2}\dk/\pi^{2}=\sqrt{2\gamma
  L^{3}}/\pi^{2}$ is the probability to find a resonance with width smaller
than $\gamma$ per unit $k$-interval. This is equivalent to
Eq.~(\ref{resodens}).

\section{}\label{app:classical}

For the \tj, the matrix elements of $\tilde M$ in Eq.~(\ref{cclass})
are the absolute squares of the elements of $\tilde W$ in Eq.~(\ref{Wtj}),
\begin{equation}\label{M0tj}
\tilde M=\begin{pmatrix}
0&0&\frac{1}{4}&\frac{1}{4}\\[1mm]
0&0&\frac{1}{4}&\frac{1}{4}\\
1&0&0&0\\
0&1&0&0
\end{pmatrix}\,.
\end{equation}
The upper right block contains the probabilities to scatter from a bond
directed inward ($1\to0$ or $2\to0$) into an outward bond ($0\to1$ or
$0\to2$).  The lower left block represents the probabilities for the opposite
process. This block is a 2$\times$2 unit matrix because along a path on the
graph the bond $0\to1$ is always followed by $1\to 0$ and the same holds for
$0\to2$, $2\to0$. Thus each path contains an even number $2t$ of directed
bonds, where the topological time $t=t_{1}+t_{2}$ counts the number of
excursions to vertex $1$ or vertex $2$. A path with topological time $t$ pics
up $t$ matrix elements 1 from the lower left and $t-1$ elements $1/4$ from the
upper right block, i.e. it has a weight $4^{1-t}$ (excluding the probabilities
to enter (leave) the interior graph at the start (end) of the path.

For $\tilde M(z)$ in Eq.~(\ref{Mz}) we find from Eq.~(\ref{M0tj}) and with the
substitution $x_{1,2}(z)=e^{-zL_{1,2}}$ 
\begin{equation}
  \tilde M(z)=\frac{1}{4}\begin{pmatrix}
0&0&x_{1}&x_{1}\\[1mm]
0&0&x_{2}&x_{2}\\
4x_{1}&0&0&0\\
0&4x_{2}&0&0
\end{pmatrix}\,.
\end{equation}
and a straightforward calculation yields 
\begin{align}\label{tjdetz}
  \det(1-\tilde M(z))&=1-\frac{x_{1}^{2}}{4}-\frac{x_{2}^{2}}{4}
  \\&=1-\frac{1}{4}e^{-2L_{1}z}-\frac{1}{4}e^{-2L_{2}z}\\
  \frac{d}{dz}\det(1-\tilde M(z))&=\frac{L_{1}}{2}e^{-2L_{1}z}+\frac{L_{2}}{2}e^{-2L_{2}z}
\end{align}
To complete the information required in Eq.~(\ref{finalresult}) we need the
adjugate of $I-\tilde M(z)$. As in the calculation of $S$ in
\ref{app:quantum} it suffices to calculate the lower left block (outward to
inward)
$$\frac{1}{4}\begin{pmatrix}
  4x_{1}-x_{1}x_{2}^{2}&x_{1}^{2}x_{2}\\
  x_{1}x_{2}^{2}&4x_{2}-x_{1}^{2}x_{2}
\end{pmatrix}\,.$$
Only for these matrix elements one the factor $|\tau_{d,0}\tau_{d',0}|^{2}$ is
non-zero and has the value 1/4. Finally the sum over $d,d'$ in
Eq.~(\ref{finalresult}) yields $(x_{1}(z)^{2}+x_{2}(z)^{2})/4=1-\det(I-\tilde M(z))$.
In particular, at a zero of the determinat this is just 1.
\end{appendix}

\section*{References}

\end{document}